\documentclass{PoS}

\usepackage[utf8]{inputenc}
\usepackage{amsmath,amssymb,graphicx,textcomp,enumerate,alltt,xspace,multirow,xcolor}
\usepackage{subfig}
\usepackage{fancyvrb}
\usepackage{mathtools}
\usepackage[numbers,sort&compress]{natbib}

\renewcommand{\thefigure}{\arabic{figure}} 

\newcommand\f[2]{\frac{#1}{#2}} 
\renewcommand\to{\rightarrow} 
\newcommand\nn{\nonumber}

\newcommand\rcut{r_{\rm cut}}

\newcommand\zmin{z_{\rm min}}
\newcommand\zmax{z_{\rm max}}
\newcommand\sigmah{{\hat \sigma}}

\title{Recent developments in $q_T$ subtraction: EW corrections and power suppressed contributions}

\ShortTitle{Recent developments in $q_T$ subtraction}

\author{\speaker{Luca Buonocore}\\
        Dipartimento di Fisica, Universit\`a di Napoli Federico II and INFN, Sezione di Napoli, I-80126 Napoli, Italy, Physik Institut, Universit\"at Z\"urich, CH-8057 Z\"urich, Switzerland\\
        E-mail: \email{luca.buonocore@na.infn.it}}

\abstract{The $q_T$ subtraction formalism represents a well established and successful technique to deal with the computation of QCD radiative corrections up to NNLO (and beyond) for a large class of processes relevant at the LHC. We have explored the possibility to apply $q_T$ subtraction to the computation of EW corrections with the (final) aim to develop a subtraction formalism suitable for the computation of mixed QCD-EW corrections. We present numerical results for the complete NLO EW corrections to the Drell-Yan production of a massive lepton pair.
Furthermore, we have investigated the structure of the power suppressed contributions at small-$q_T$ in this process and we present new analytical results on the effects of the soft radiation emitted off a charged massive final state.}
\FullConference{14th International Symposium on Radiative Corrections (RADCOR2019)\\ 
9-13 September 2019\\
		Palais des Papes, Avignon, France.\\
                Based on arXiv:1911.10166}

\begin{document}

\section{Introduction}

The $q_T$ subtraction formalism \cite{Catani:2007vq} is a method to handle and cancel the InfraRed (IR) divergences associated to the computation of higher order QCD radiative corrections. It has been originally formulated to deal with the hadroproduction of color singlet systems and applied successfully to a variety of such processes up to next-to-next-to the leading order (NNLO) (see Ref.~\cite{Grazzini:2017mhc} and references therein).
A first application to the approximate computation of N$^3$LO corrections to Higgs boson production through gluon fusion have been recently presented~\cite{Cieri:2018oms}. Thanks to the progress in the formulation of transverse momentum resummation for heavy quark pair production~\cite{Zhu:2012ts,Li:2013mia,Catani:2014qha,Angeles-Martinez:2018mqh}, $q_T$ subtraction has been recently extended to the computation of this process up to NNLO~\cite{Bonciani:2015sha,Catani:2019iny,Catani:2019hip}. So far the method has never been applied to the computation of electroweak (EW) corrections. 

We recall the differential $q_T$ subtraction formula for the inclusive reaction $h_1 + h_2 \to F + X $, $F$ being either a color singlet system or a heavy-quark pair, in hadron-hadron collisions~\cite{Catani:2007vq}.
\begin{equation}
\label{eq:main}
d{\sigmah}^{F}_{(N)NLO}={\cal H}^{F}_{(N)NLO}\otimes d{\sigmah}^{F}_{LO}
+\left[ d{\sigmah}^{F+\rm{jet}}_{(N)LO}-
d{\sigmah}^{F, \, CT}_{(N)NLO}\right],
\end{equation}
where $d{\sigmah}^{F+\rm{jet}}_{(N)LO}$ is the $F$+jet cross 
section at (N)LO accuracy. The square bracket term of Eq.~(\ref{eq:main}) is IR finite in the limit $q_T \to 0$, but its individual contributions,
$d{\sigmah}^{F+\rm{jet}}_{(N)LO}$ and the auxiliary subtraction cross section $d{\sigmah}^{F, \, CT}_{(N)NLO}$, are separately divergent. The IR counterterm is built by exploiting the infrared behavior of the transverse-momentum distribution $q_T$ of $F$ and, hence, by construction the $q_T$ counterterm is non-local~\cite{Catani:2007vq}

In practice, the integration is performed introducing a lower cut-off $\rcut$ on the transverse momentum of the produced system $q_T$ divided by its mass $M$
\begin{equation}
  \label{eq:idef}
    \overline{I} \equiv \int \left[d\sigma_\text{LO}^{F+\text{jet}}-d\sigma^\text{CT}\right] \rightarrow I(r_\text{cut}) \equiv \int \left[d\sigma_\text{LO}^{F+\text{jet}}-d\sigma^\text{CT}\right] \Theta\left(\frac{q_T}{M}-r_\text{cut}\right).
\end{equation}
Thanks to this slicing prescription, the integral is finite and the two contributions can then be integrated separetely. They develop logarithmic singularities in $\rcut$, which globally cancel among each other, plus spurius terms vanishing in the limit $\rcut \to 0$. Hence, a residual $\rcut$-dependence, in the form of power suppressed terms (modulo logarithmic enhancements), is introduced in the computation, 
\begin{equation}
  I(r_\text{cut}) = \overline{I} + {\cal{O}}(\rcut^n).
\end{equation}
The efficiency of the subtraction procedure crucially depends on the size of such power suppressed contributions.
For the inclusive color singlet production, the leading power suppressed contribution has been investigated~\cite{Grazzini:2016ctr} and explicitly computed at NLO~\cite{Ebert:2018gsn,Cieri:2019tfv}, finding a quadratic behavior ($n=2$). The presence of fiducial cuts may worsen the $\rcut$ dependence leading in some case to a linear behavior~\cite{Ebert:2019zkb}. In the case of heavy-quark production the $\rcut$ dependence is found to be {\it linear} \cite{Catani:2017tuc,Catani:2019iny,Catani:2019hip}, and it is an interesting question to investigate the origin of this peculiar behavior.

\noindent In this talk, we report our results on 
\begin{itemize}
  \item the first application of the $q_T$ subtraction formalism to the computation of NLO EW corrections. The formulation of the method for heavy-quark production can indeed be straightforwardly extended to the computation of EW corrections to the Drell-Yan process by means of a well-established abelianisation procedure. This is a first step towards the development of a suitable subtraction scheme for mixed QCD-EW and NNLO EW corrections;
  \item the analytical computation of the power suppressed terms to the $q_T$ subtraction fomula in the presence of soft radiation off a massive final-state at NLO. We discuss a procedure to incorporate fiducial cuts as well.   
\end{itemize}

\section{NLO EW corrections to the Drell-Yan process}
\label{sec:nloew}

\subsection{Implementation}
The structure of the subtraction has been derived exploiting the abelianisation
procedure outlined in Ref.~\cite{deFlorian:2018wcj}. 
We have implemented our calculation as an extension of the numerical program of Ref.~\cite{Catani:2009sm}.
All the required tree level matrix elements are computed analytically while the virtual EW corrections for $q{\bar q}\to  l^+ l^-$, which include vertex and box diagrams, are obtained by using {\sc Gosam} \cite{Cullen:2011ac,Cullen:2014yla}
. To set the notation, we denote with $\sigma_{LO}^{q{\bar q}}$ and $\sigma^{\gamma\gamma}_{LO}$ the Born level cross sections in the $q{\bar q}$ and $\gamma\gamma$ channels, respectively.
At NLO EW we have
\begin{equation}
  \sigma_{NLO}=\sigma_{LO}^{q{\bar q}}+\sigma^{\gamma\gamma}_{LO}+\Delta\sigma_{q{\bar q}}+\Delta\sigma_{q\gamma}+\Delta\sigma_{\gamma\gamma}
\end{equation}
where we have introduced the ${\cal O}(\alpha^3)$ correction in the $q{\bar q}$ channel, $\Delta\sigma_{q{\bar q}}$, the corresponding correction in the $q({\bar q})\gamma$ channel, $\Delta\sigma_{q\gamma}$,
and the correction in the $\gamma\gamma$ channel, $\Delta\sigma_{\gamma\gamma}$. Since the $\gamma\gamma$ channel provides only a very small contribution to the Drell-Yan cross section, $\Delta\sigma_{\gamma\gamma}$
will be neglected in the following discussion.

\subsection{Numerical validation}
To validate our implementation, we have repeated our calculation by using the dipole subtraction method \cite{Catani:2002hc} and the independent matrix-element generator {\sc Recola}~\cite{Actis:2012qn,Actis:2016mpe} for the virtual corrections.

We use the setup of Ref.~\cite{Dittmaier:2009cr}, and, in particular, we work in the $G_\mu$ scheme \footnote{More precisely, real and virtual photons emissions are controlled by $\alpha(0)$, while the $\alpha^2$ in the LO cross section is derived from $G_F$, $m_Z$ and $m_W$.}
with
\begin{align}
  & G_F=1.16637\times 10^{-5}~{\rm GeV}^{-2}  & \alpha(0)=1/137.03599911\\
  & m_W=80.403~{\rm GeV}                     & m_Z=91.1876~{\rm GeV}\\
  & \Gamma_W=2.141~{\rm GeV}                 & \Gamma_Z=2.4952~{\rm GeV}
\end{align}
and use the complex-mass scheme~\cite{Denner:2005fg} throughout.
Following Ref.~\cite{Dittmaier:2009cr}, the MRST2004qed \cite{Martin:2004dh} parton distribution functions (PDFs) are used.
In order to avoid large logarithmic terms in the lepton mass which may complicate the numerical convergence
we set the mass of the final-state lepton to $m_l=10$ GeV.
\renewcommand{\arraystretch}{1.6}
\begin{table}[h]
  \centering
\begin{tabular}{ccc}
  \hline
  & $q_T+\text{GoSam}$  &  {\sc cs}+{\sc Recola}\\
  \hline
  $\sigma_{LO}^{q\bar{q}}$ (pb) & 
   \multicolumn{2}{c}{$683.53 \pm 0.03$} \\
  \hline
  $\Delta\sigma_{q\overline{q}}$ (pb) & $-5.920 \pm 0.034$ &  $-5.919 \pm 0.008$ \\
  \hline
  $\sigma_{LO}^{\gamma\gamma}$ (pb) & 
   \multicolumn{2}{c}{$ 1.1524 \pm 0.0004 $} \\
  \hline
  $\Delta\sigma_{q\gamma}$ (pb) & $-0.6694 \pm 0.0008$ & $-0.6690 \pm 0.0005$ \\
  \hline
\end{tabular}
\caption{\label{tab:cmp_deltaqq} \small Comparison of NLO EW corrections to the Drell-Yan process computed with $q_T$ subtraction and dipole subtraction. In the $q{\bar q}$ channel the $q_T$ result is obtained with a linear extrapolation in the $\rcut\to 0$ limit (see Figure~\ref{figs:rcut-var-EW}),
while in the $q({\bar q})\gamma$ channel it is obtained at $\rcut = 0.01 \%$. The LO result in the $q{\bar q}$ and $\gamma\gamma$ channels is also reported for reference.}
\end{table}
\noindent The following set of cuts are applied
\begin{equation}\label{eq:cuts}
  m_{ll}>50~{\rm GeV}~~~~~~~~~~p_{T,l}>25~{\rm GeV}~~~~~~~~~~~~|y_l|<2.5\, .
\end{equation}

We report our result in Table~\ref{tab:cmp_deltaqq}. The NLO correction $\Delta\sigma_{q{\bar q}}$ is obtained performing the calculation at different values of $\rcut$ and extrapolating to $\rcut\to 0$ through a linear fit.
Our results are compared with the corresponding results obtained with dipole subtraction (CS+{\sc Recola}).
We see that the two results are in perfect agreement.

\subsection{Dependence on $\rcut$}
\begin{figure}[h]
  \centering
  \includegraphics[width=0.48\textwidth]{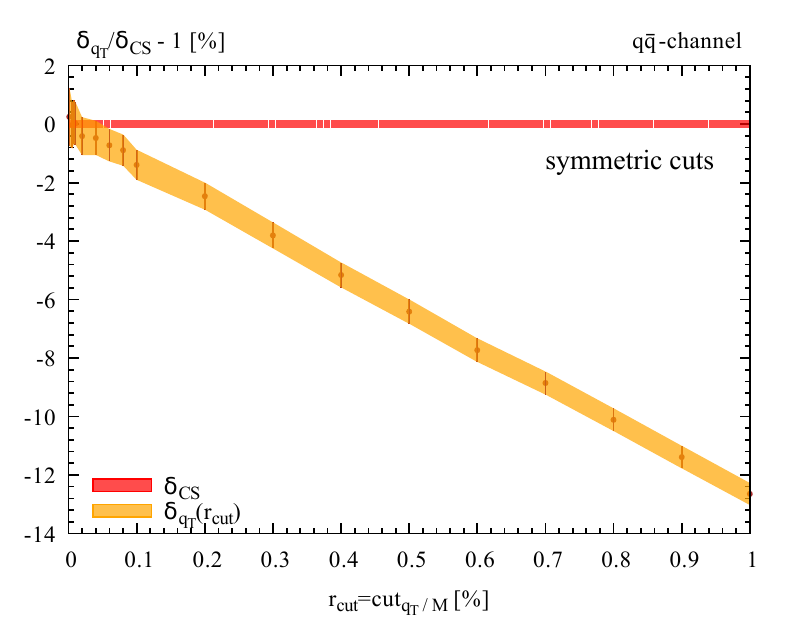}
  \includegraphics[width=0.48\textwidth]{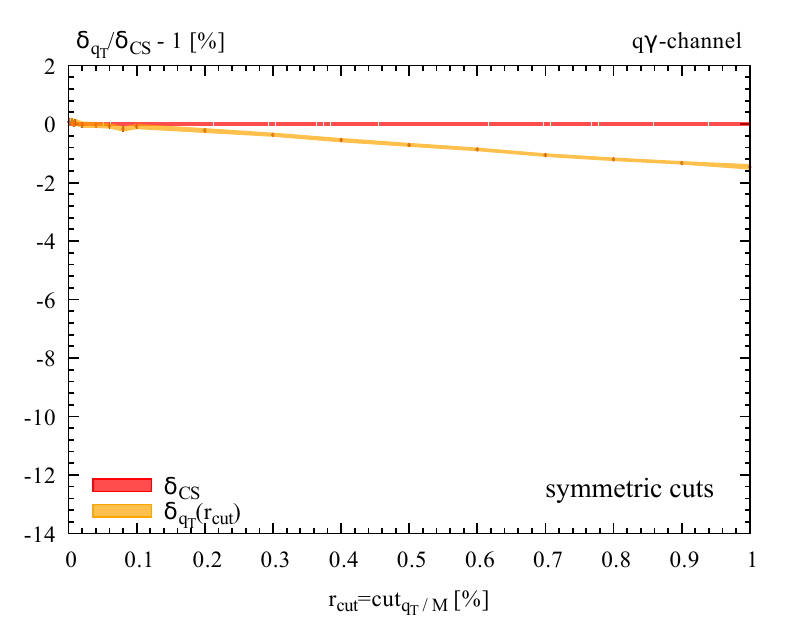}
  \caption{\label{figs:rcut-var-EW} \small NLO EW correction as a function of $\rcut$ in the dominant $q\overline{q}$ diagonal channel (left panel) and in the off-diagonal $q({\bar q})\gamma$ channel (right panel) at $14\,$TeV. The NLO result is normalised to the $\rcut$-independent cross section computed with dipole subtraction. The lepton mass is fixed to $m_l=10\,$GeV. The fiducial cuts in Eq.~\eqref{eq:cuts} are applied.}
\end{figure}

\begin{figure}
  \centering
  \includegraphics[width=0.48\textwidth]{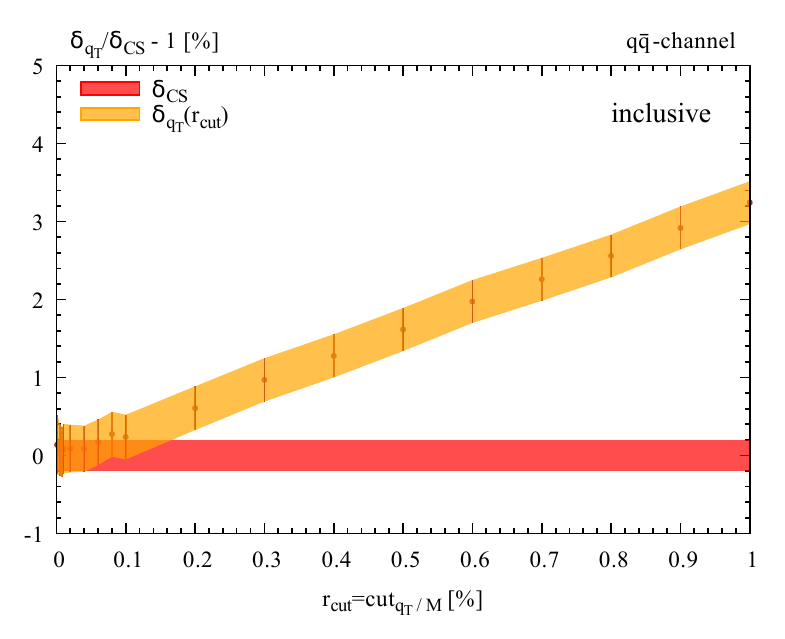}
  \includegraphics[width=0.48\textwidth]{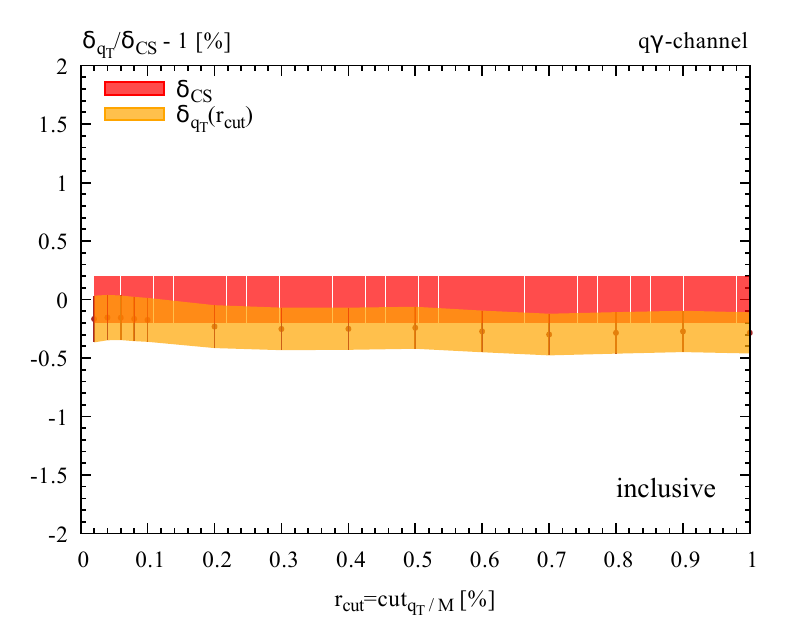}
  \caption{\label{figs:rcut-var-EW-nocuts} \small NLO EW correction as a function of $\rcut$ in the dominant $q\overline{q}$ diagonal channel (left panel) and in the off-diagonal $q({\bar q})\gamma$ channel (right panel) at $14\,$TeV. The NLO result is normalised to the $\rcut$-independent cross section computed with dipole subtraction. The lepton mass is fixed to $m_l=10\,$GeV. No cuts are applied.}
\end{figure}

We have varied $\rcut$ in the range $0.01\% \leq \rcut \leq 1\%$ and we have used the $\rcut$-independent cross section computed with our inhouse implementation of the dipole subtraction method as normalisation.
The results for the $\rcut$ dependent correction $\delta_{q_T}=\Delta\sigma/\sigma_{LO}^{q{\bar q}}$ in the $q{\bar q}$ and $q\gamma$ channels are shown in Figure~\ref{figs:rcut-var-EW}. A distictive linear behavior in the dominant 
$q\bar{q}$-annihilation channel emerges. Nonetheless 
it is known that symmetric cuts on the transverse momenta of the final 
state leptons challenge the convergence of $q_T$-subtraction leading 
to a stronger dependence on $\rcut$ even in the case in which a charge-neutral final state is produced~\cite{Grazzini:2017mhc,Ebert:2019zkb}.
In Figure~\ref{figs:rcut-var-EW-nocuts} we show the dependence of the NLO 
corrections for the inclusive cross section on $\rcut$ when no cuts 
are applied. Again a distinct linear behavior in the dominant 
$q\bar{q}$-annihilation channel emerges, in agreement with what has already 
been observed for the case of the $t{\bar t}$ cross section \cite{Catani:2019iny}, which 
can be clearly interpreted as a genuine new effect due to the 
emission of radiation off the massive final-state leptons.

\subsection{Physical lepton masses}

Before concluding this section, we briefly show that the method outlined above can be effectively employed
for realistic phenomenological studies pushing the lepton mass to the physical muon value.

Indeed, the lepton mass acts as a regulator of final state collinear divergences and within the 
$q_T$ subtraction formalism it is required to be kept to a finite value.
While there are not any issues from the formal point of view, small values of the lepton mass may
challenge the numerical implementation spoiling the accuracy of the method.

We have repeated our calculation by setting the mass of the heavy lepton
to the physical muon mass  $m_l\equiv m_\mu=105.658369\,$MeV, using the setup and fiducial cuts as before.
To have a fully independent validation we have compared our results with those obtained with the well established 
public generator {\sc Sanc} \cite{Andonov:2004hi}.
Our results are reported in Table~\ref{tab:cmp_sanc}, and show that the agreement is quite good, at few {\it per mille} on the NLO correction.

\renewcommand{\arraystretch}{1.6}
\begin{table}[h]
\centering
\begin{tabular}{ccc}
  \hline
  & $q_T+\text{GoSam}$  & {\sc Sanc}\\
  \hline
  $\Delta\sigma_{q\overline{q}}+\Delta\sigma_{q\gamma}$ (pb) & $-29.95 \pm 0.04$ &  $-29.99 \pm 0.02$ \\
  \hline
\end{tabular}
\caption{\label{tab:cmp_sanc} \small Tuned comparison for NLO EW corrections to the Drell-Yan process with $m_l = m_\mu=105.658369\,$ MeV with the {\sc Sanc} generator. The $q_T$ result is the limiting value for $\rcut\to 0$ obtained with a linear fit for the NLO correction in the diagonal $q\bar{q}$-annihilation channel, and it is the value at $\rcut = 0.01 \%$ for the off-diagonal $q({\bar q})\gamma$ channel.}
\end{table}

\section{Power corrections}
\label{sec:power}

We study the behavior of NLO inclusive cross sections computed with $q_T$ subtraction in the $\rcut\to 0$ limit with the aim to determine the structure of the leading power correction and the origin of the observed linear behavior.
We focus on the production of a massive lepton pair in pure QED in the diagonal channel
  \begin{equation}
    q(p_1)+{\bar q}(p_2)\to l^+(p_3) l^-(p_4)+\gamma(k)
    \end{equation}
  with $p_3^2=p_4^2=m^2$.
In the following, we sketch the strategy of the computation and present the main results. For further details we refer to~\cite{Buonocore:2019puv}. 

\subsection{Structure of the computation}
The power suppressed terms arise by the integration of real emission cross section and the counterterm, Eq.~\eqref{eq:idef}. Since we keep $\rcut$ finite, we can consider their contributions separately. The small-$\rcut$ behavior of the counterterm is well understood~\cite{Bozzi:2005wk} and leads to quadratic power corrections.
Therefore, we focus on the real emission cross section, which must be the source of the observed linear behavior, and we further split it into three gauge-invariant contributions: final-state radiation, initial-state radiation and their interference. 
\noindent The $\rcut$ dependence of the total partonic cross section can be recasted in the following formula
\begin{equation}
  \label{eq:master}
    \frac{d \sigmah_{q\bar q}}{d r_{\text{cut}}^2} =-\frac{1}{32}\frac{1}{(2\pi)^4} \int_{\zmin}^{\zmax} \frac{z\, dz}{\sqrt{(1 -
        z)^2 - 4 z \rcut^2}} \sqrt{1 - \frac{\zmin}{z}} \int d \Omega | \mathcal{M} |^2 
  \end{equation}
where $\Omega$ is the solid angle of the dilepton in the frame in which the the two leptons are back-to-back and  
\begin{equation}\label{eq:zlimits}
  \zmin=\f{4m^2}{s}~~~~~~~~~~~~~\zmax=1-2\rcut\sqrt{1+\rcut^2}+2\rcut^2\, .
\end{equation}
The soft limit corresponds to $z\to1$. The presence of the cutoff prevents the upper integration to reach the singular point. In the small-$\rcut$, indeed, it behaves as $\zmax\sim 1- 2\rcut$, and it is linear with respect to $\rcut$.

\noindent Hence, our procedure consists in  
\begin{enumerate}
\item integrating the matrix element over the angle $d\Omega$;
\item extracting the small $\rcut$ behavior up to the desidered order by means of a uniform expansion with respect to the $z$ variable.    
\end{enumerate}
The integral in $z$ is indeed curbersome to be computed analytically while, on the other hand, we do not need to know the exact result for our purpose.

The interference contribution is odd under the exchange $p_3\leftrightarrow p_4$ and therefore vanishes after angular integration. Therefore we focus on final- and initial-state only. 

Eq.~\eqref{eq:master} contains already the main features relevant to understand the $\rcut$ dependence. We observe that both the upper integration limit $\zmax$ and the integrand function depend on $\rcut$. After performing the angular integration, the matrix element leads to an analytic function of $\rcut^2$ (at fixed $z$), while, regarded as function of $z$, it is divergent in the soft limit $z\to1$. The universal jacobian square root factor in Eq.~\eqref{eq:master} is function of $\rcut^2$ and vanishes at $z=\zmax$. As mentioned above, the upper limit behaves linearly and one naively may expect that it would control the overall behavior by power counting.  

\noindent Our finding is that:
\begin{itemize}
\item for final-state radiation, the angular average of the matrix element as function of $z$ behaves as $\f{1}{(1-z)^2}$ in the soft limit, as dictated by the the universal eikonal approximation. This eventually leads to a linear behavior in $\rcut$.
\item for initial-state radiation, the angular average of the matrix element as function of $z$ is finite in the soft limit. The vanishing of the jacobian factor on the upper limit changes the power counting from linear to quadratic.
\end{itemize}

\subsection{Results}
\label{sec:results}
\noindent We pass now to present the results obtained for the final- and initial-state contributions. Their expressions are more coveniently written in terms of the variable $\beta=\sqrt{1-\f{4m^2}{s}}$, the common lepton velocity in the partonic center-of-mass frame. In the following, we denote with $\sigma_0(s)$ the  Born cross section 
\begin{equation}
  \sigma_0(s)=\f{2\pi}{9s}\alpha^2 e_q^2\beta(3-\beta^2)
\end{equation}
The $\rcut$ dependence of the partonic cross section is 
\begin{itemize}
\item {\it Final-state radiation}
\begin{equation}\label{eq:final_result_fsr}
\begin{split}
  \sigmah^{\rm FS}_{q\bar q}(s;r_\text{cut})&=\sigma_0(s)\f{\alpha}{2\pi}\bigg\{\left[2-\frac{(1+\beta^2)}{\beta}\ln\frac{1+\beta}{1-\beta}\right] \ln{(\rcut^2)}\\
  &- \frac{3\pi}{8} \left[ \frac{6(5-\beta^2)}{3-\beta^2} + \frac{-47+8\beta^2+3\beta^4}{\beta(3-\beta^2)}\ln\frac{1+\beta}{1-\beta} \right]r_\text{cut} \bigg \}+ O(r_\text{cut}^2)\\
&\equiv \sigmah^{\rm FS}_{\rm LP}(s;\rcut) + \sigmah^{\rm FS}_{\rm NLP}(s;\rcut) + O(r_\text{cut}^2)
\end{split}
\end{equation}
where we have dropped terms which do not depend on $\rcut$.
Eq.~(\ref{eq:final_result_fsr}) shows that the next-to-leading power contribution $\sigmah^{\rm FS}_{\rm NLP}(s;\rcut)$ is linear in $\rcut$ and it is responsible for the behavior observed in Figure~\ref{figs:rcut-var-EW-nocuts}.
\item {\it Initial-state radiation}
\begin{align}
\label{eq:final_result_isr}
~~~~\sigmah^{\rm IS}_{q\bar q}(s;r_\text{cut})&= \sigma_0(s)\frac{\alpha}{2\pi}e_q^2 \bigg\{ \ln^2{\rcut^2}- 4 \left( 2\ln{2} - \frac{4}{3} - \ln\frac{1-\beta^2}{\beta^2} - \frac{1}{\beta(3-\beta^2)}\ln\frac{1+\beta}{1-\beta} \right) \ln{\rcut^2}\nn\\ 
  & -\frac{3}{2}\frac{(1+\beta^2)(1-\beta^2)^2}{\beta^4(3-\beta^2)} \left(1-4\ln{2} +2\ln\frac{(1-\beta^2)\rcut}{\beta^2}\right) \rcut^2 \bigg\} +..........\nn\\
  &\equiv \sigmah^{\rm IS}_{\rm LP}(s;\rcut) + \sigmah^{\rm IS}_{\rm NLP}(s;\rcut) +..........
\end{align}
where we have dropped terms which do not depend on $\rcut$ and the dots stand for terms that vanish faster than $\rcut^2$ as $\rcut\to 0$.
As expected, the next-to-leading power contribution $\sigmah^{\rm IS}_{\rm NLP}(s;\rcut)$ is quadratic in $\rcut$, modulo logarithmic enhancements.
\end{itemize}

\section{Final-state radiation at next-to-leading power: beyond inclusive observables}
\label{sec:diff}

The results presented in Sec.\ref{sec:results} have been obtained for the most inclusive observable, the total partonic cross section. This has been a crucial point in order to perform the computation analytically. 
In what follows we outline a strategy to remove the final-state linear power suppressed contribution from the $q_T$ subtration formula at NLO at differential level.
Let us start from eq.~\eqref{eq:main} at NLO
\begin{equation}
\label{eq:main2}
d{\sigmah}^{F}_{NLO}={\cal H}^{F}_{NLO}\otimes d{\sigmah}^{F}_{LO}
+\left[ d{\sigmah}^{F+\rm{jet}}_{LO}-
d{\sigmah}^{F, \, CT}_{NLO}\right]\Theta\left(\f{q_T}{M}-\rcut \right),
\end{equation}
where we have written explicitly the $\rcut$ constraint. Consider the following extension
\begin{equation}
\label{eq:main3}
d{\sigmah}^{F}_{NLO}={\cal H}^{F}_{NLO}\otimes d{\sigmah}^{F}_{LO}
+\left[ d{\sigmah}^{F+\rm{jet}}_{LO}-
d{\sigmah}^{F, \, CT}_{NLO}\right]\Theta\left(\f{q_T}{M}-\rcut \right) + \left[ d{\sigmah}^{F+\rm{jet}}_{FS,LO}-
d{\sigmah}^{F, \, CT}_{S,NLO}\right]\Theta\left(\rcut - \f{q_T}{M} \right).
\end{equation}
In the above formula, $d{\sigmah}^{F+\rm{jet}}_{FS,LO}$ is the differential cross section associated to final-state radiation only and $d{\sigmah}^{F, \, CT}_{S,NLO}$ is an arbitrary counterterm which cancels the corresponding final-state soft divergence, so that their difference is finite. The new term lives in the unresolved region below $\rcut$ so that its contribution is vanishing in the limit $\rcut\to0$. Thus, the first observation is that Eq.\eqref{eq:main2} and Eq.\eqref{eq:main3} differ only by power suppressed terms in $\rcut$, and therefore formally equivalent.
The second and key observation is that, if the new counterterm, as the standard $q_T$ subtraction one, does not introduce additional linear power corrections, then the subtraction formula in Eq~\eqref{eq:main3} is free of linear power suppressed terms. Indeed, the linear term arising from the real emission cross section exactly cancels in the sum
\begin{equation}
\left[ d{\sigmah}^{F+\rm{jet}}_{LO}-
d{\sigmah}^{F, \, CT}_{NLO}\right]\Theta\left(\f{q_T}{M}-\rcut \right) + \left[ d{\sigmah}^{F+\rm{jet}}_{FS,LO}-
d{\sigmah}^{F, \, CT}_{S,NLO}\right]\Theta\left(\rcut - \f{q_T}{M} \right) = {\bar I} + {\cal O}(\rcut^2).
\end{equation}
The above condition is fulfilled if the soft counterterm is built as the product of the eikonal approximation for the matrix element times the soft phase space. 
The third and last observation is that if the soft subtraction is local, then it is effectively possible to perform the integration in the unresolved region. 

Therefore, to construct the additional soft counterterm we only need a soft mapping which reabsorbs the radiation into a Born-like configuration.  
Among the available mappings at NLO, we choose the mapping proposed in Ref.~\cite{Buonocore:2017lry}.
As in the standard FKS mapping~\cite{Frixione:1995ms}, we identify an emitter and a radiated parton. The radiation variables are given by the radiation energy fraction 
$\xi = 2 E_\text{rad}/\sqrt{s}$ ($s$ is the partonic CM energy), the cosine of the angle between the emitter and the radiated partons $y$
and an azimuthal angle $\phi$ (we refer to Ref.~\cite{Buonocore:2017lry} for more details). 
The phase space reads
\begin{equation}
  d\Phi_R = d\Phi_B \times J(\xi,y,\phi) d\xi dy d\phi
\end{equation}
where $d\Phi_B$ is the Born phase space element and the jacobian $J$ is given in Eq.~(49) of Ref.~\cite{Buonocore:2017lry}, and reduces to $J^{(0)}=s\,\xi/(4\pi)^3$ in the soft limit, which is what we actually need.
Then, we define the local soft counterterm as
\begin{equation}
  \label{eq:localsoftct}
  d \sigmah^{CT}_{S} = d\sigmah_{LO}(\Phi_B) \times \frac{e^2}{4\pi^3s}\frac{d\xi}{\xi}dy d\phi\left[\frac{s-2m^2}{(1-\beta y_\text{phy})(1+\beta y_\text{phy})} -\frac{m^2}{(1-\beta y_\text{phy})^2} - \frac{m^2}{(1+\beta y_\text{phy})^2}\right]
\end{equation} 
where $\beta = \sqrt{1-{4m^2}/{s}}$ and $y_\text{phy}$ is the cosine of the physical angle between the emitted photon and the leptons in the Born configuration (in practice we have either $y_\text{phy}=y$ or $y_\text{phy}=-y$ \cite{Buonocore:2017lry}).  

\begin{figure}[htb]
    \includegraphics[width=0.48\textwidth]{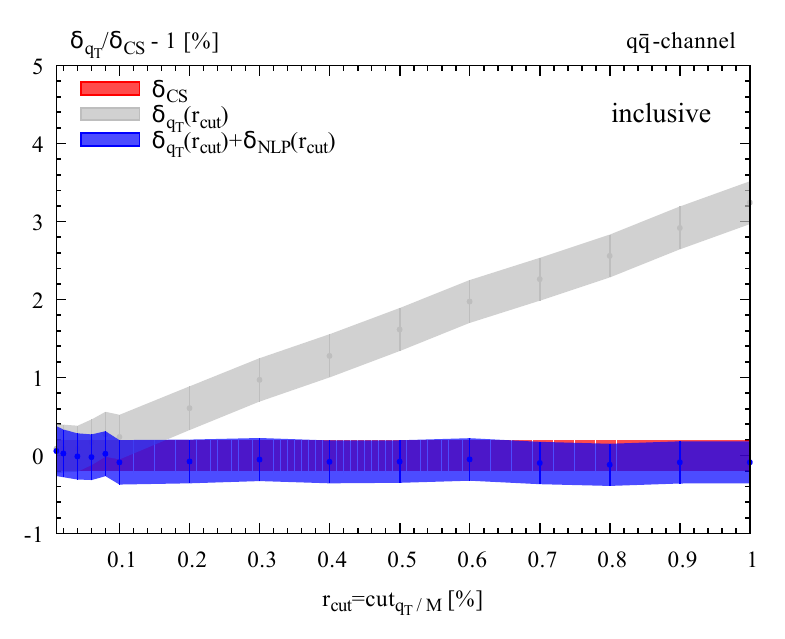}
    \includegraphics[width=0.48\textwidth]{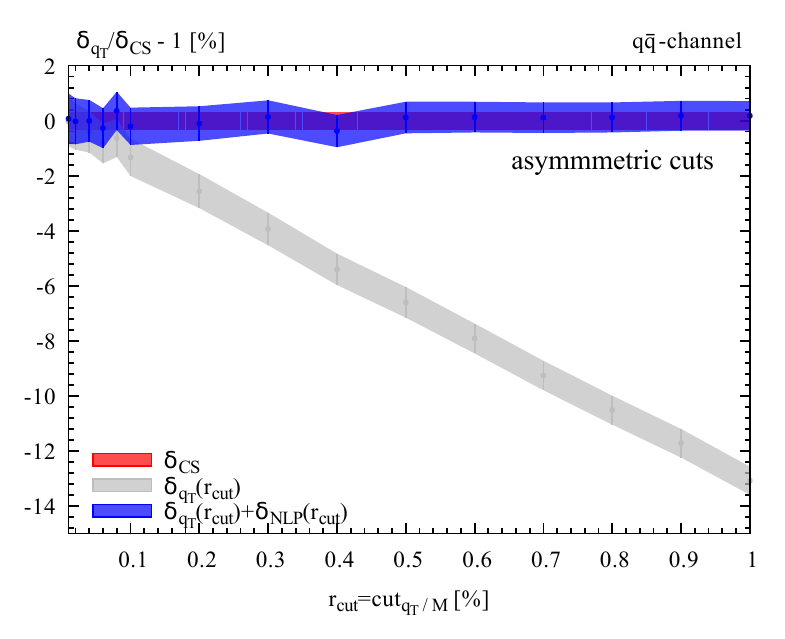}
    \caption{\label{fig:gamma+Z_lincorr}
      \small NLO EW correction as a function of $\rcut$ for the complete Drell-Yan process in the dominant $q\overline{q}$ diagonal channel without cuts (left panel) and with asymmetric cuts (right panel) at $7\,$TeV. The standard result obtained with $q_T$ subtraction (grey band) is compared with the result obtained by including the power suppressed contribution according to the modified subtraction formula in Eq.~(\ref{eq:main3}).
      The NLO result is normalised to the $\rcut$-independent cross section computed with dipole subtraction.}
\end{figure}
 
In Figure~\ref{fig:gamma+Z_lincorr} we study the $\rcut$ dependence of the NLO EW correction to the complete Drell-Yan process applying the ``improved'' $q_T$ subtraction formula in Eq.~\eqref{eq:main3}. We consider $pp$ collisions at $\sqrt{S}=7$ TeV and we compute the $\rcut$ dependent correction $\delta_{q_T}(\rcut)$ in the case in which no cuts are applied (Figure~\ref{fig:gamma+Z_lincorr} (left)) and when asymmetric cuts on the transverse momenta and rapidities are applied: $p_{T,l^-}>25\,$GeV, $p_{T,l^+}>20\,$GeV and $|y_l|<2.5$ (Figure~\ref{fig:gamma+Z_lincorr} (right)).
We see that in both cases the linear dependence with $\rcut$ is nicely cancelled\footnote{As discussed in Sec.~\ref{sec:nloew}, when symmetric cuts are applied a linear dependence on $\rcut$ appears in the contribution from initial-state radiation.}.

\section{Conclusions}

In this talk, we have presented the first application of the $q_T$ subtraction formalism to the computation of EW radiative corrections, namely the NLO EW corrections to the hadroproduction of a massive dilepton system through the Drell-Yan mechanism. Our calculation paves the way to possible applications to the computation of mixed QCD-EW corrections \cite{Dittmaier:2014qza,Dittmaier:2015rxo,deFlorian:2018wcj,Delto:2019ewv,Bonciani:2019nuy} and to NNLO QED corrections \cite{deFlorian:2018wcj} to the Drell-Yan process. 

We then have discussed the power suppressed terms in the $\rcut$ regulator which affect the $q_T$ subtraction formula when also the radiation off massive final state is taken into account. Specializing to the case of hadroproduction of a dilepton system, we have reported explicit analytical expressions for the power corrections to the inclusive partonic cross section in the diagonal $q{\bar q}$ channel, proving that final-state radiation is responsible for the change in the power correction from quadratic to linear. We have presented a procedure to ``improve'' the NLO $q_T$ subtraction formula removing the linear power corrections at fully differential level. 

\bibliographystyle{apsrev4-1}
\bibliography{biblioqt}

\end{document}